\documentclass{article}
\usepackage{spconf,amsmath,graphicx}
\usepackage{amsmath, amssymb}
\usepackage{subcaption}
\usepackage{titlesec}

\titlespacing*{\section}
{0pt}{8pt plus 1pt minus 1pt}{8pt plus 1pt minus 1pt}

\titleformat{\section}
  {\large\bfseries}
  {\thesection}{1em}{} 

\titleformat{\subsection}
  {\normalsize\bfseries} 
  {\thesubsection}{1em}{} 

\title{Cross-Modal Synthesis of Structural MRI and Functional Connectivity Networks via Conditional ViT-GANs}
\name{Yuda Bi, Anees Abrol, Jing Sui, and Vince Calhoun}
\address{Tri-institutional Center for Translational Research in Neuroimaging and Data Science @\{GSU, GATech, Emory\}}
\begin{document}

%
\maketitle
\begin{abstract}
The cross-modal synthesis between structural magnetic resonance imaging (sMRI) and functional network connectivity (FNC) is a relatively unexplored area in medical imaging, especially with respect to schizophrenia. This study employs conditional Vision Transformer Generative Adversarial Networks (cViT-GANs) to generate FNC data based on sMRI inputs. After training on a comprehensive dataset that included both individuals with schizophrenia and healthy control subjects, our cViT-GAN model effectively synthesized the FNC matrix for each subject, and then formed a group difference FNC matrix, obtaining a Pearson correlation of 0.73 with the actual FNC matrix. In addition, our FNC visualization results demonstrate significant correlations in particular subcortical brain regions, highlighting the model's capability of capturing detailed structural-functional associations. This performance distinguishes our model from conditional CNN-based GAN alternatives such as Pix2Pix. Our research is one of the first attempts to link sMRI and FNC synthesis, setting it apart from other cross-modal studies that concentrate on T1- and T2-weighted MR images or the fusion of MRI and CT scans.
\end{abstract}
\section{Introduction}
\label{sec:intro}

Generative adversarial networks (GANs) originated as a novel approach to create generative models using a generator and discriminator trained together \cite{goodfellow2014generative}. GANs have revolutionized image generation, style adaptation, and data augmentation \cite{radford2015unsupervised, isola2017image}. The use of modality translation tasks in medical imaging is especially notable \cite{zhan2021multi}. The vision transformer (ViT) has revolutionized visual tasks by modeling long-range dependencies and offering a reliable alternative to CNNs by utilizing language processing architectures \cite{dosovitskiy2020image}. The ViT-GAN combines ViT's characteristics with GAN, achieving superior image synthesis performance without CNN components \cite{lee2021vitgan, jiang2021transgan}. However, the application of ViT-GANs to medical datasets, particularly brain MRI, remains largely unexplored.

This study presents the conditional ViT-GAN (cViT-GAN), a new generative framework for synthesizing images from sMRI to functional network connectivity data (FNC), which utilizes sMRI and class identifier as conditions and is regulated by newly designed correlation loss. FNC expresses temporal correlations between independent neural activities and is commonly depicted as a 2D matrix. To derive FNC matrices, independent component analysis (ICA) was applied to fMRI datasets from the same cohort as the sMRI samples \cite{du2020neuromark}. Existing literature highlights the intricate relationship between structural and functional MRI modalities \cite{uddin2013complex, rahaman2021multi}, which are often complementary in nature. Early research indicates that the integration of sMRI and fMRI could enhance diagnostic accuracy for disorders such as schizophrenia \cite{bi2022multicrossvit}. Although fMRI data are abundant in temporal information, analyzing them can be computationally demanding. FNC serves as a more efficient alternative, summarizing complex temporal patterns into a 2D layout without significant diagnostic loss. Our research aims to comprehend the intricate correlation between sMRI and FNC in diagnosing and understanding schizophrenia. Meanwhile, our cViT-GAN model also showed promising FNC visualization. The model created FNC matrices that showed functional zones identical to the original FNC data using sMRI data. This research illuminates the pathogenesis of schizophrenia by pinpointing these functional areas.

\begin{figure*}[htb]
    \centering
    \includegraphics[width=17cm]{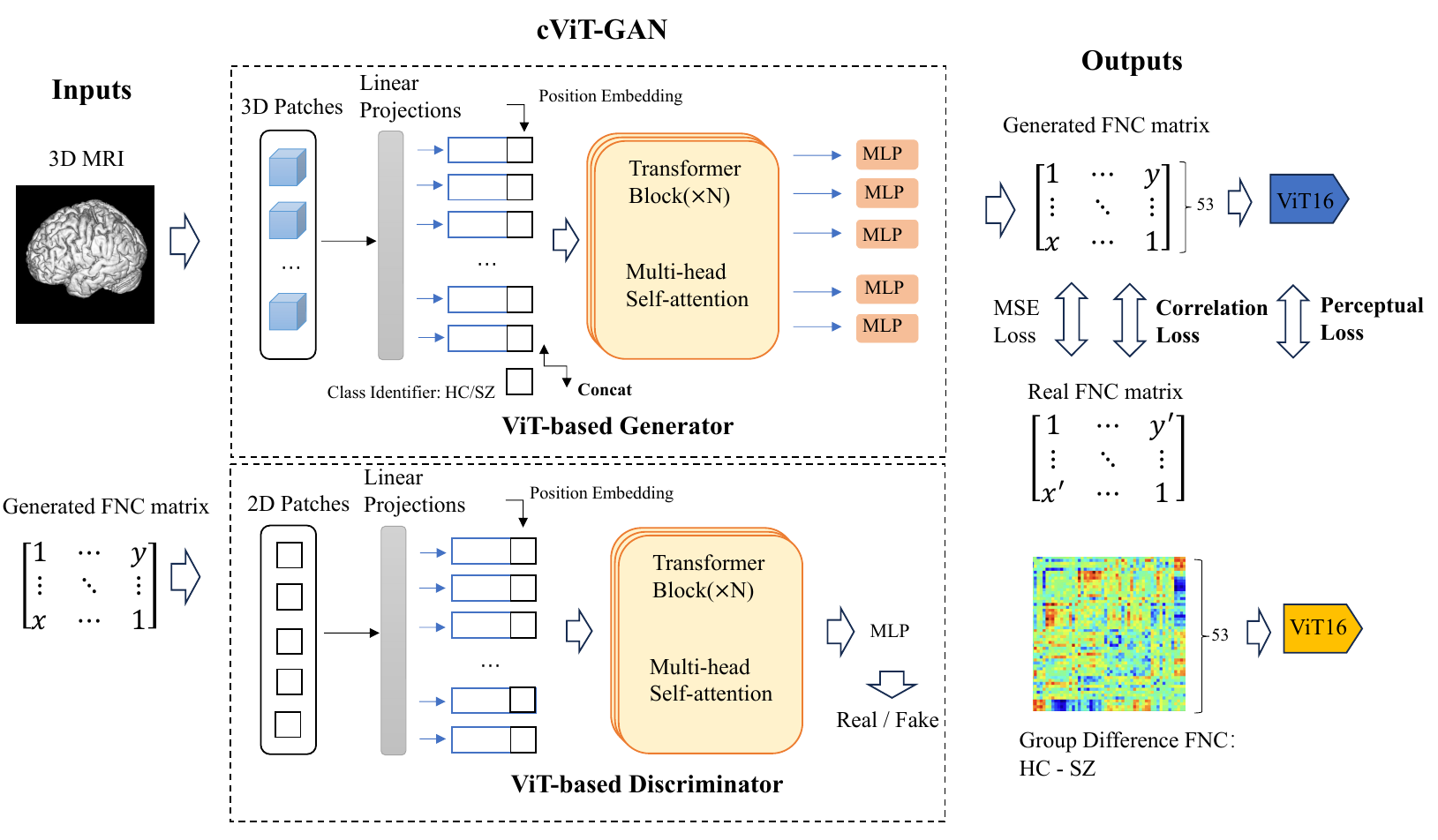}
    \caption{The cViT-GAN framework with the embedded class identifier for sMRI to FNC synthesis: The model is regularized by VGG perceptual loss and correlation loss. }
    \label{fig:my_label}
\end{figure*}

\section{Related Works}
\label{sec:rw}

Recent studies have applied GANs and transformer-based architectures to medical imaging tasks, showing superior performance in image synthesis and reconstruction \cite{nie2018medical, cole2020unsupervised, wang2021dicyc, pan2022transformer}. Specifically, Pan et al. used a transformer-based encoder for translating MRI modalities, demonstrating enhanced performance \cite{pan2022transformer}. Li et al. introduced MedViTGAN for generating synthetic histopathology images, incorporating an auxiliary classifier and adaptive loss mechanism \cite{li2022medvitgan}. Additionally, Abrol et al. showed that fusing multi-feature sMRI and fMRI data via deep learning significantly improves the prediction of Alzheimer's disease progression \cite{abrol2019multimodal}.

\section{Methods}
\label{sec:methods}

In this section, we explain the methods employed in our cViT-GAN model for image synthesis, specifically for generating FNC matrices from sMRI inputs. We detail the generator and discriminator architectures, as well as the composite loss functions designed to achieve highly accurate FNC matrix synthesis. Importantly, our model incorporates multiple loss components to obtain more accurate results.

\subsection{Conditional ViTGAN}

The cViT-GAN consists of two main components: a generator \( G \) and a discriminator \( D \). The architecture leverages the principles of GAN to optimize the following objective function:
\begin{align}
\min_{G} \max_{D} \mathcal{L}(G, D) &= \mathbb{E}_{y \sim p_{\text{data}}(y)}[\log D(y)] \nonumber \\
&+ \mathbb{E}_{x \sim p_{\text{input}}(x)}[\log(1 - D(G(x)))]
\end{align}

In the objective function \(\mathcal{L}(G, D)\), \( \mathbb{E}_{y \sim p_{\text{data}}(y)}[\log D(y)] \) represents the expectation of the log-likelihood that the discriminator \( D \) correctly classifies real FNC matrices \( y \). The term \( \mathbb{E}_{x \sim p_{\text{input}}(x)}[\log(1 - D(G(x)))] \) reflects the expectation that \( D \) misclassifies the FNC matrices \( \hat{y} = G(x) \) generated from sMRI inputs \( x \).

\textbf{Generator}: The generator's purpose is to transform 3D sMRI inputs \( x \) into a 2D network-by-network Functional Network Connectivity (FNC) matrix \( \hat{y} \). This conversion is accomplished through a series of intricate transformer blocks, making use of the inherent patterns and features of the sMRI data.

\textit{Position Embedding with Class Identifier}: Before processing through the generator, the 3D sMRI undergoes a preliminary transformation. It is divided into a set of 3D patches, which are subsequently mapped to linear space to derive position embedding vectors. The introduction of a class identifier is crucial as it provides the model context about the nature of the input – whether it's from a schizophrenia patient or a healthy control.

\textit{Transformer Block}: The heart of the generator is the transformer block. Here, the structure captures the complexities of the data through a combination of multi-head self-attention and position-wise feed-forward neural networks. The sequential nature of this design allows the generator to capture long-range dependencies in the data. The formulas show the computation:
\begin{align}
Z &= \text{LayerNorm}(\text{SelfAttention}(X) + X) \\
\text{Output} &= \text{LayerNorm}(Z + \text{FFN}(Z))
\end{align}

\textit{MLP for FNC Fragments}: Post the transformer blocks, each resultant embedding vector undergoes further refinement through a multi-layer perceptron (MLP). This step generates fragments, smaller pieces, of the desired FNC matrix. The final stage involves stitching these fragments together seamlessly, giving us the synthesized FNC matrix \( \hat{y} \).

\textbf{Discriminator}: In any generative setup, the discriminator \( D \) plays a pivotal role. In this architecture, \( D \) is trained to discern between authentic FNC matrices \( y \) and the ones generated by our model \( \hat{y} \). Its design draws inspiration from the ViT model, which has established itself as a potent tool for classification tasks in the realm of images.

\subsection{Loss Functions}

The standard GAN loss may not be sufficient for capturing the intricate relationships in FNC matrices. Therefore, our model employs a composite loss function, which incorporates several other loss components.

\textbf{ViT Perceptual Loss:} The ViT Perceptual loss aims to capture high-level features and patterns in FNC matrices, moving beyond mere pixel-level differences. Instead of relying on the hierarchical structure of CNN like VGG, we utilize the attention mechanisms of a pre-trained ViT-16 network to calculate this loss between real and generated FNC matrices. The ViT Perceptual loss is defined as:

\begin{equation}
\mathcal{L}_{\text{ViT\_16}} = \frac{1}{W_l H_l} \sum_{i,j} \left( F^l(y)_{ij} - F^l(\hat{y})_{ij} \right)^2
\end{equation}

Here, \( F^l(\cdot) \) is the feature map extracted from the \( l \)-th block of the ViT-16 network. \( W_l \) and \( H_l \) are the width and height, respectively, of the feature map at block \( l \).

\textbf{Correlation Loss:} The correlation loss captures relationships between corresponding regions in the FNC matrix. This is particularly important as FNC matrices often show similarity not just in absolute values but also in the relative arrangement of values. The correlation loss is defined as:

\begin{equation}
\mathcal{L}_{\text{corr}} = 1 - \frac{\text{cov}(y, \hat{y})}{\sigma_y \sigma_{\hat{y}}}
\end{equation}

In this equation, \( \text{cov}(y, \hat{y}) \) is the covariance between \( y \) and \( \hat{y} \), and \( \sigma_y \) and \( \sigma_{\hat{y}} \) are their respective standard deviations.

\textbf{Total Losses:} In total, the loss function of our cViT-GAN is:
\begin{equation}
\mathcal{L}_{G} = \mathcal{L}_{\text{GAN}} + \lambda_1 \mathcal{L}_{\text{MSE}} + \lambda_2 \mathcal{L}_{\text{ViT\_16}} + \lambda_3 \mathcal{L}_{\text{corr}}
\end{equation}

Here, \( \mathcal{L}_{\text{GAN}} \) is the GAN loss for the generator. The hyperparameters \( \lambda_1 \), \( \lambda_2 \), and \( \lambda_3 \) control the balance of MSE loss, VIT perceptual loss, and correlation loss, respectively.

\section{Experiments}
\label{sec:exp}

\begin{table*}[ht]
\centering
\caption{Comparison of Pearson correlation and cosine similarity of cViT-GAN and baselines.}
\label{table:model-comparison}
\begin{tabular}{c|c|c|c|c|c|c}
\hline
Model & Generator & Discriminator & Class Identifier & Correlation Loss & Pearson & Cosine \\
\hline
cViT-GAN & ViT & ViT & Yes & Yes & \textbf{0.731} & \textbf{0.732} \\
\hline
DCGAN & U-Net & MLP & Yes & Yes & 0.695 & 0.693 \\
\hline
Pix2Pix \cite{isola2017image} & U-Net & PatchGAN & Yes & Yes & 0.719 & 0.714 \\
\hline
cViT-GAN-0 & ViT & ViT & No & Yes & 0.543 & 0.544 \\
\hline
cViT-GAN-1 & ViT & ViT & Yes & No & 0.724 & 0.722 \\
\hline
\end{tabular}
\end{table*}

\begin{figure*}
    \centering
    \begin{subfigure}{0.38\textwidth} 
        \centering
        \includegraphics[width=0.8\textwidth]{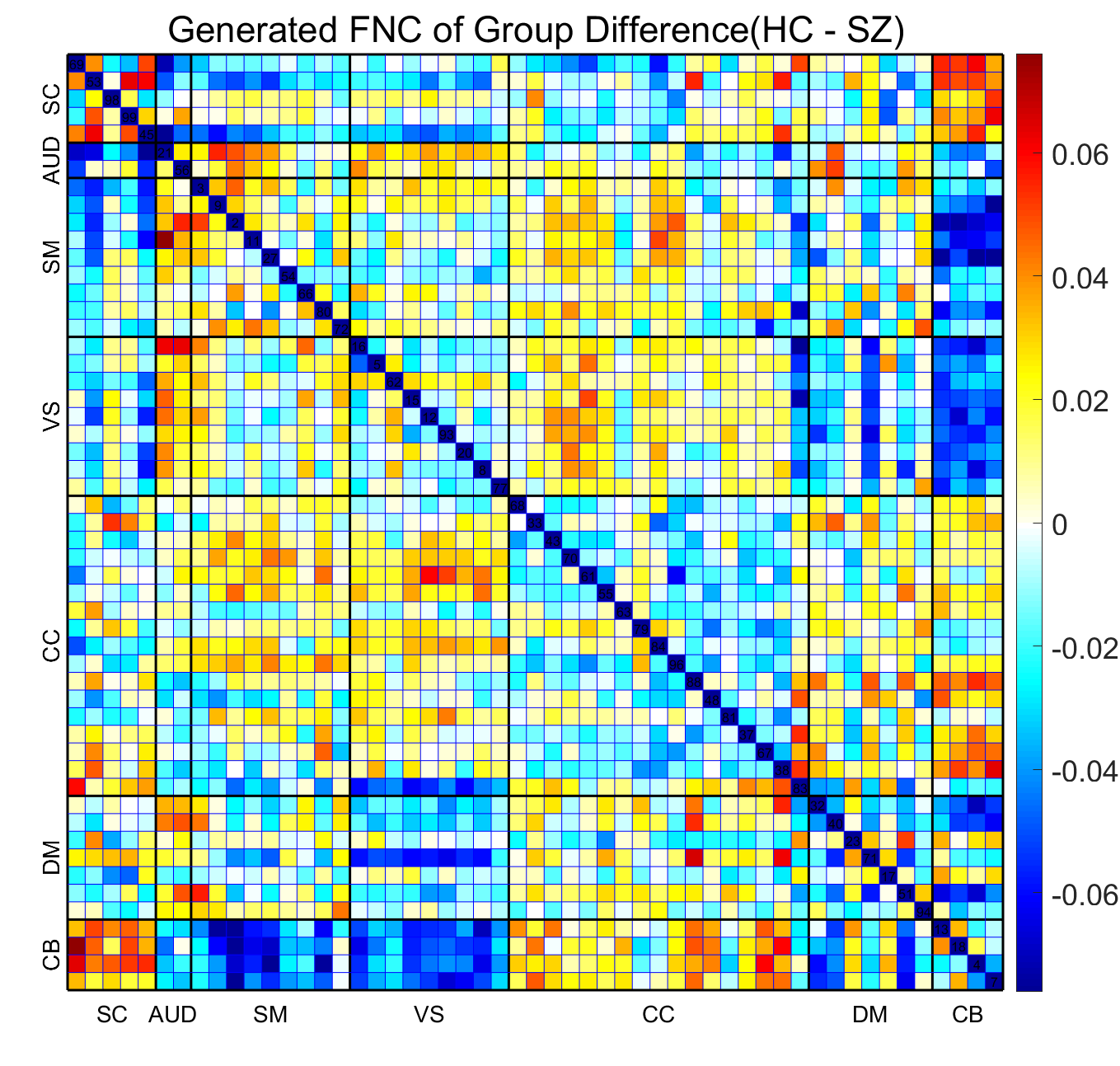} 
        \caption{}
        \label{fig:subfig-a}
    \end{subfigure}
    \begin{subfigure}{0.38\textwidth} 
        \centering
        \includegraphics[width=0.8\textwidth]{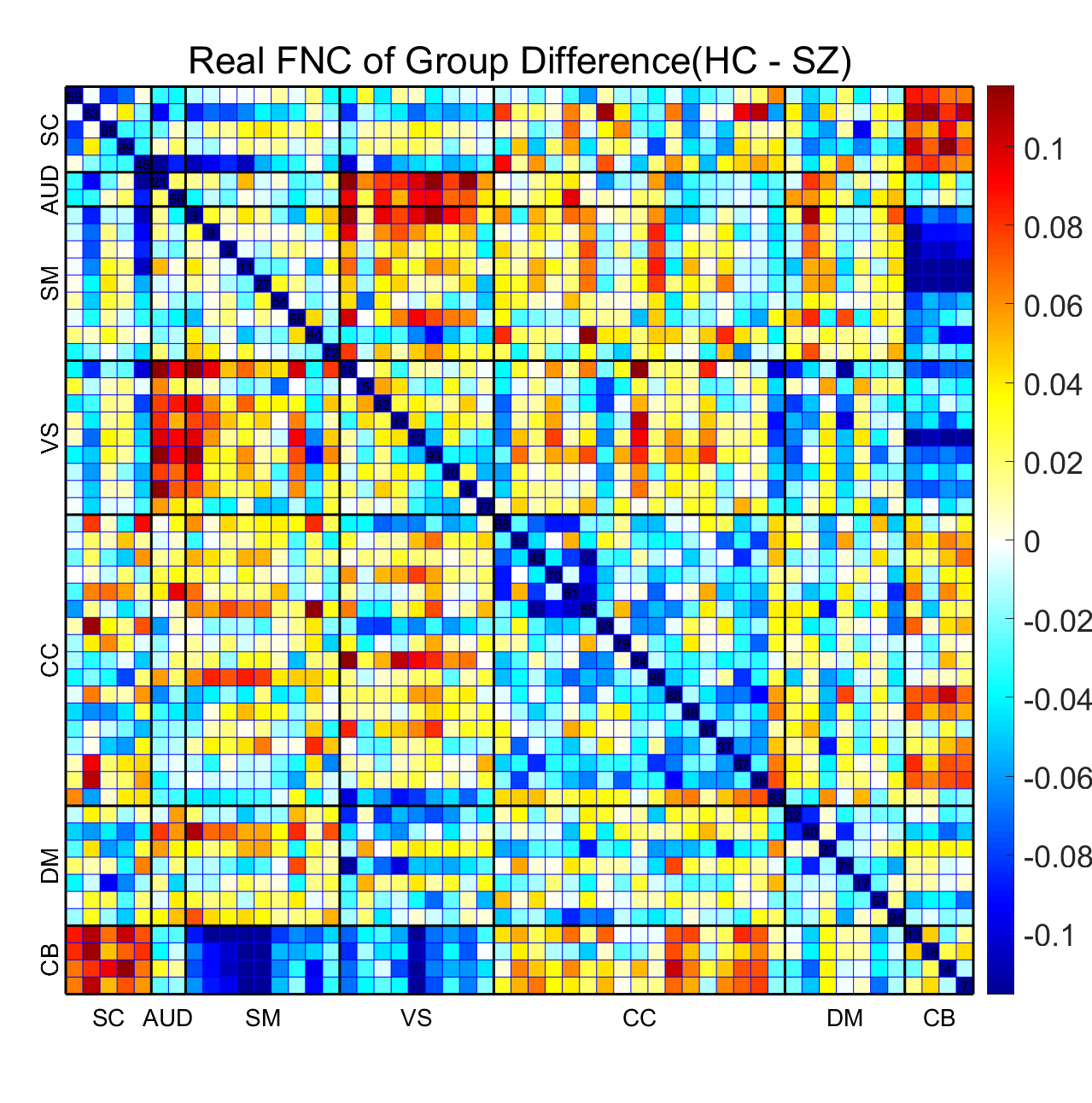} 
        \caption{}
        \label{fig:subfig-b}
    \end{subfigure}
    \caption{Generated FNC (a) and real FNC (b) for the group difference analysis of schizophrenia. }
    \label{fig:main}
\end{figure*}

\textbf{Datasets}: We employed two datasets related to clinical research on schizophrenia, sourced from multiple sites in the United States and China, featuring a total of 827 and 815 participants, respectively. The first dataset amalgamated data from three studies—fBIRN, MPRC, and COBRE—using nine different 3-Tesla scanners with standard echo planar imaging (EPI) sequences. The second dataset originated in China, obtained using three different 3-Tesla scanners. Both datasets comprised a mix of control individuals and schizophrenia patients and included detailed demographic characteristics such as mean age and gender distribution.

\textbf{Preprocessing}: The sMRI data were preprocessed with a pipeline consisting of gray-scale segmentation and normalization. We enhanced the robustness of our model by introducing data augmentation techniques, including random rotation and Gaussian noise addition. For the fMRI data, preprocessing involved slice timing correction, realignment, and normalization. The FNC and sMRI feature vectors were derived following methods from our previous research \cite{du2020neuromark}. Specifically, we employed independent component analysis (ICA) to extract time courses, which were subsequently used to compute the FNC matrices via cross-correlation.

\textbf{Models}: In our experiment, we deployed the cViT-GAN model, consisting of both a generator and a discriminator shown in Figure 1. For comparative analysis, we included the following baseline models: \textbf{1)DCGAN}: A deep convolutional generative adversarial network (DCGAN) was employed, where both the generator and the discriminator are built using CNNs. \textbf{2)Pix2Pix} \cite{isola2017image}: This model utilizes a U-Net architecture for the generator and a PatchGAN architecture for the discriminator, focusing primarily on image-to-image translation tasks. \textbf{3)cViT-GAN-0}: In this variation of the cViT-GAN model, we only used sMRI as the condition for generating outputs, without including any class identifiers during concatenation. \textbf{4)cViT-GAN-1}: This model is another variant of cViT-GAN. It uses both sMRI and class identifiers for conditionality. However, this version of the model excludes correlation loss and VGG perceptual loss during the training phase.

\textbf{Training and Evaluation}: 
All models were trained using the AdamW optimizer. The initial learning rate was set uniformly to \(1 \times 10^{-4}\) across all architectures, including cViT-GAN, its variations, DCGAN, and Pix2Pix. The training lasted for 300 epochs and was executed on 8 NVIDIA V100 GPUs in a distributed environment. Both the generator and discriminator used a MultiStepLR scheduler, with the learning rate being decreased by a factor of 0.1 at epoch 50 and again at epoch 150. To assess the models' performance comprehensively, we employed a 10-fold cross-validation approach. For evaluation purposes, we compared the generated FNC matrices against actual FNC matrices produced by each model. Our evaluation metrics comprised Pearson correlation and cosine similarity in order to capture both linear and angular relationships between the real and generated matrices. These metrics offer a thorough comprehension of the extent to which the produced FNC matrices emulate the structural and functional traits existing in the real data.

\section{Results}
\label{sec:results}

In this section, we compare the experimental results of our conditional cViT-GAN model with various baseline models, specifically focusing on generating ``group difference FNC" between the healthy control group (HC) and schizophrenia patients (SZ). We employ Pearson correlation and cosine similarity as our evaluation metrics for these generative methods. Beyond quantitative measures, we also spotlight functional domains where our generated group difference FNC closely matches the actual FNC. This high degree of comparability may underscore significant structural-functional relationships between these groups.

\textbf{Basic results}: Table 1 provides a comprehensive comparison of the Pearson correlation and cosine similarity scores among various generative models, including our proposed cViT-GAN. In terms of Pearson correlation and cosine similarity, the cViT-GAN outperforms all other baseline models, achieving a score of \(0.731\) and \(0.732\) respectively. This highlights the advantage of using a ViT architecture for both the generator and discriminator along with class identifiers and the newly designed correlation loss. 

\textbf{FNC domain visualizations}: 
This analysis provides visual representations of group difference functional network connectivity matrices (HC-SZ). Figure 2 shows the comparison between the generated group difference FNC with the real one. Remarkably, our cViT-GAN model is proficient at generating FNC matrices that closely mimic actual group difference FNC data, specifically in subcortical regions such as CB-SC, CB-AUD, CB-SM, CB-VS, CB-CC, CB-DM, and CB-CB. In these critical subcortical areas, the correlation between synthetic FNC derived from sMRI and actual group difference FNC data reaches an impressive similarity of up to 0.85. This pivotal finding not only underscores the importance of subcortical structures in distinguishing between HC and SZ groups but also validates the accuracy and utility of our cViT-GAN model in capturing these essential structural-functional connections. The high degree of similarity in subcortical regions suggests that our cViT-GAN model is exceptionally adept at reproducing complex, real-world neurological patterns, especially those that mark differences between healthy and pathological states. This offers promising avenues for advancing our understanding of disorders like schizophrenia, equipping us with more accurate diagnostic tools and targeted treatment strategies. Moreover, our model also revealed significant similarities in differential values for other region pairs, including SC-SM, CC-CC, SM-DM, and VS-DM, with moderate similarities noted in pairs like VS-AUD and CC-SM. These observations contribute additional layers of understanding to how differences in functional networks between the HC and SZ groups might be influenced by foundational structural abnormalities. Collectively, this knowledge is pivotal for the development of improved diagnostic methods and treatment plans that can effectively tackle the complexities of conditions like schizophrenia.

\section{Conclusion}
\label{sec:conclusion}

In conclusion, our study reveals that sMRI data can closely reflect FNC, offering the potential for one imaging modality to inform the other in diagnostics. This deepens our understanding of the brain's structure-function relationship, paving the way for personalized medicine and disease prediction. Our work sets the stage for holistic biomarkers through integrated neuroimaging, and the generative model enables FNC simulation under specific pathological states, yielding new insights into brain functionality and behavior.

\vfill\pagebreak

\bibliographystyle{IEEEbib}
\bibliography{strings,refs}

\end{document}